\documentclass[12pt]{article}
\usepackage{amssymb,amsfonts,stmaryrd}
\begin{document}

\begin{titlepage}

                            \begin{center}
                            \vspace*{2cm}
        \large\bf Algebra \ and \ calculus \ for \ Tsallis \ thermostatistics\\

                            \vfill

              \normalsize\sf    NIKOS \ \ KALOGEROPOULOS\\

                            \vspace{0.2cm}

 \normalsize\sf Department of Science\\
 BMCC - The City University of New York,\\
 199 Chambers St., \ New York, NY 10007, \ USA\\

                            \end{center}

                            \vfill

                     \centerline{\normalsize\bf Abstract}
\normalsize\rm\setlength{\baselineskip}{24pt}

\noindent We construct generalized additions and multiplications,
forming  fields, and division algebras inspired by  the Tsallis
thermo-statistics. We also construct  derivations and integrations
in this spirit. These operations do not reduce to the naively
expected ones, when the deformation parameter approaches zero.

                             \vfill


                             \vfill

\noindent\rule{8cm}{0.2mm}\\
\begin{tabular}{ll}
\small\rm E-mail: & \small\rm nkalogeropoulos@bmcc.cuny.edu\\
                  & \small\rm nkaloger@yahoo.com
\end{tabular}
\end{titlepage}


                            \newpage

\normalsize\rm\setlength{\baselineskip}{24pt}

\centerline{\sc 1. \ Introduction}

                            \vspace{3mm}

Non-extensive thermo-statistics has attracted a lot of attention
in recent years. This is due, in part, to the  contributions of C.
Tsallis and his collaborators who have been  advocating the use of
the non-extensive entropy
\begin{equation}
          S_q = k_B \ \frac{1-\sum\limits_{i\in W}p_i^q}{q-1}
\end{equation}
associated with a probability distribution \ $p_i, \ i\in W$, \
where \ $W$ \ is the set of cells in which one divides the phase
space of a system in a coarse grained description, and $k_B$ is
the Boltzmann constant. Applications of this form of entropy have
been found in very diverse areas ranging from Dynamical Systems
theory and Physics to Medicine, Linguistics and Social Sciences
[2]. Other forms of entropy have also been postulated and
advocated over the years, (see [2] and references therein for some
definitions) claiming to provide generalizations of the
Boltzmann-Gibbs entropy to a variety of systems. In all these
cases someone recovers the Boltzmann-Gibbs entropy in the limit of
deformation disappearing \ $q\rightarrow 1$. \ Someone can notice
that the Tsallis entropy (1) for two probabilistically independent
systems \ $A$ \ and \ $B$, \ obeys
\begin{equation}
  S_q(A+B) = S_q(A) + S_q(B) + (1-q) S_q(A) S_q(B)
\end{equation}
as opposed to the usual addition \ $S_1(A+B) = S_1(A)+S_1(B)$ \
that the Boltzmann-Gibbs entropy follows. The name ``non-extensive
entropy" is a result of (2). The generalized additivity (2) forces
us to review the conventional definition of additivity [3] and to
construct a generalized algebraic and analytic framework which
will express such ideas more naturally [4]-[12]. The present work
follows the spirit of [7], [8], [11] and modifies as well as
extends the work of [12] and [13].\\

A brief summary of the contents of this paper is as follows: In
Section 2 we introduce the \ $\stackrel{k}{\bigcirc}$ \ operations
of generalized addition and multiplication, and based on them, we
construct a vector space and a division algebra. In Section 3, we
introduce the \ $\stackrel{k}{\square}$ \ operations and construct
similar algebraic structures as in Section 2. In Section 4, we
construct generalized derivatives and integrals corresponding to
these operations. In Section 5, we make some
general comments and point to some topics for future research.\\

                                  \vspace{5mm}




\centerline{\sc 2. \ The \ $\stackrel{k}{\bigcirc}$ \ algebraic \
operations}

                                \vspace{3mm}

There is, a priori, an infinity of ways in which we can define a
generalized addition and a generalized multiplication. We narrow
down our choices by the requirement that the sought after
operations should reflect, as much as possible, the algebraic
properties of Tsallis' entropy. Let \ $k\in\mathbb{R}$ \ indicate
the non-extensive parameter [7], [8], [10], [11]. Other authors
use \ $1-q$, \ $q-1$ \ and \ $\alpha$ \ instead of \ $k$ \
[1]-[6], [9], [12], [13].\\

\noindent We start by using (2.1) of [11] with the \ $k$-deformed
logarithm \ $\ln_k(x)$
\begin{equation}
 x_{\{k\}} = \ln_k(x) :=\frac{x^k-1}{k}
\end{equation}
and consequently (2.2) of [11] with the \ $k$-deformed exponential
\ $e_k(x)$
\begin{equation}
 x^{\{k\}} = e_k(x) := (1+kx)^\frac{1}{k}
\end{equation}
With this identification, the requirements (2.3) and (2.4) of [11]
are satisfied if we define for \ $x,y \in\mathbb{R}$ \ the
generalized addition
\begin{equation}
  x\stackrel{k}{\oplus} y  \ = \ (x^k+y^k-1)^\frac{1}{k}
\end{equation}
We see that the \ $\stackrel{k}{\oplus}$ \  addition reduces to
the usual multiplication in \ $\mathbb{R}$ \ as \ $k\rightarrow
0$, \ namely
\begin{displaymath}
            \lim_{k\rightarrow 0} \ x\stackrel{k}{\oplus}y \ = \ xy
\end{displaymath}
 We can verify that \ $\stackrel{k}{\oplus}$ \ is
commutative, associative and has \ $1$ \ as neutral element. The
opposite of \ $x$, \ denoted by \ $\stackrel{k}{\ominus}x$, \  is
given by
\begin{equation}
   \stackrel{k}{\ominus} x \ = \ (2-x^k)^\frac{1}{k}
\end{equation}
Subtraction  is defined as \ $x\stackrel{k}{\ominus}y =
x\stackrel{k}{\oplus}(\stackrel{k}{\ominus}y)$ \ and (5),(6) give
\begin{equation}
  x\stackrel{k}{\ominus} y \ = \ (x^k-y^k+1)^\frac{1}{k}
\end{equation}
The generalized multiplication is defined by
\begin{equation}
 x\stackrel{k}{\otimes} y \ = \ \left\{\frac{(xy)^k - x^k - y^k
 +(k+1)}{k}\right\}^\frac{1}{k}
\end{equation}
We notice that
\begin{displaymath}
  \lim_{k\rightarrow 0} \ x\stackrel{k}{\otimes} y \ = \
   e^{(\ln x)(\ln y)}
\end{displaymath}
 We see that \
$\stackrel{k}{\otimes}$ \ is commutative, associative and has
identity element \ $(k+1)^\frac{1}{k}$. In addition, we observe
that the distributivity property holds between \
$\stackrel{k}{\oplus}$ \ and \ $\stackrel{k}{\otimes}$, \ namely
\begin{equation}
  x\stackrel{k}{\otimes}(y\stackrel{k}{\oplus}z) \ = \
   (x\stackrel{k}{\otimes}y) \stackrel{k}{\oplus}
   (x\stackrel{k}{\otimes}z)
\end{equation}
Therefore, the structure $\mathcal{R}_1 = (\mathbb{R},
\stackrel{k}{\oplus}, \stackrel{k}{\otimes})$ is a commutative
ring with identity [14]. We can prove, by induction, that by
$\stackrel{k}{\oplus}$-adding $n\in\mathbb{Z}_+$ times $x$,
\begin{equation}
x\stackrel{k}{\oplus}\cdots\stackrel{k}{\oplus} x \ = \
 \left\{ nx^k - (n-1) \right\}^\frac{1}{k}
\end{equation}
By using (10) we check that \ $\mathcal{R}_1$ \ has zero
characteristic. Motivated by (10), we define the multiplication \
$\stackrel{k}{\odot}$, \ for \ $n\in\mathbb{R}$
\begin{equation}
    n\stackrel{k}{\odot} x \ = \
\left\{ nx^k - (n-1) \right\}^\frac{1}{k}
\end{equation}
We can prove, by induction, that by \
$\stackrel{k}{\otimes}$-multiplying \ $x$ \ by itself \
$n\in\mathbb{Z}_+$ \ times, one gets
\begin{equation}
 x\stackrel{k}{\otimes} \cdots \stackrel{k}{\otimes} x \ = \
  \left\{ \frac{(x^k-1)^n + k ^{n-1}}{k^{n-1}}\right\}^\frac{1}{k}
\end{equation}
which can be used to prove that \ $\mathcal{R}_1$ \ has no
non-trivial nilpotent elements. The inverse element of \
$x\in\mathbb{R} \setminus \{1 \} $ \ denoted by \
$\stackrel{k}{\oslash}x$ \ is
\begin{equation}
  \stackrel{k}{\oslash}x \ = \ \left\{ 1 + \frac{k^2}{x^k
  -1}\right\}^\frac{1}{k}
\end{equation}
It is natural to define [11]-[13] the division as \
$x\stackrel{k}{\oslash}y =
x\stackrel{k}{\otimes}(\stackrel{k}{\oslash}y)$ \ and find, by
combining (8) and (13),
\begin{equation}
 x\stackrel{k}{\oslash} y \ = \ \left\{ k\frac{x^k-1}{y^k-1} + 1
 \right\}^\frac{1}{k}
\end{equation}
So \ $\mathcal{R}_1$ \ is actually a field [14]. It can also be
checked that (2.6) and (2.7) of [11] become in this case
\begin{equation}
 e_k(x)\stackrel{k}{\oplus} e_k(y) \ = \ e_k(x+y)
\end{equation}
\begin{equation}
 e_k(x)\stackrel{k}{\otimes} e_k(y) \ = \ e_k(xy)
\end{equation}
respectively. To proceed, we use (11) to form an algebra
$\mathcal{A}$ \ over \ $(\mathbb{R}, +, \cdot )$. \  As sets \
$\mathcal{A} = \mathcal{R}_1$. \ Let \ $r,s\in \mathbb{R}$ \ and \
$x,y \in\mathcal{R}_1$.\ We readily check that
\begin{equation}
  r\stackrel{k}{\odot} (x\stackrel{k}{\oplus}y) \ = \
  (r\stackrel{k}{\odot} x) \stackrel{k}{\oplus}
  (r\stackrel{k}{\odot} y)
\end{equation}
\begin{equation}
  (r+s)\stackrel{k}{\odot} x \ = \
  (r\stackrel{k}{\odot}x)\stackrel{k}{\oplus}(s\stackrel{k}{\odot}x)
\end{equation}
\begin{equation}
  r\stackrel{k}{\odot}(s\stackrel{k}{\odot}x) \ = \
  (rs)\stackrel{k}{\odot} x
\end{equation}
\begin{equation}
 1\stackrel{k}{\odot}x \ = \ x
\end{equation}
We see therefore that the structure \ $\mathcal{V}_1  =
(\mathbb{R}, \stackrel{k}{\oplus}, \stackrel{k}{\odot})$ \ is a
vector space over \ $(\mathbb{R}, +, \cdot)$. \ In addition, we
find
\begin{equation}
 r\stackrel{k}{\odot}(x\stackrel{k}{\otimes}y) \ = \
 (r\stackrel{k}{\odot}x)\stackrel{k}{\otimes}y \ = \
 x \stackrel{k}{\otimes} (r\stackrel{k}{\odot}y)
\end{equation}
which means that \ $\mathcal{V}_1$ \ becomes a commutative algebra
\ $\mathcal{A}$ \ over \ $\mathbb{R}$, \ and since \
$\mathcal{R}_1$ \ is a field, \ $\mathcal{A}$ \ is actually a
commutative division algebra. It is obvious that
\begin{equation}
\dim_{\mathbb{R}}\mathcal{A} = 1
\end{equation}
Let \ $\mathbb{R}[x]$ \ denote the ring of polynomials of one
variable, with coefficients in \ $\mathbb{R}$. \ All elements of \
$\mathcal{A}$ \ are algebraic over \ $\mathbb{R}$, \ namely they
are roots of some polynomial in $\mathbb{R}[x]$. Therefore \
$\mathcal{A}$ \ is an algebraic algebra over \ $\mathbb{R}$. \
Then a theorem of Frobenius [14] and (22) imply that \
$\mathcal{A}$ \ is isomorphic to \ $\mathbb{R}$ \ or to \
$\mathbb{C}$ \ as division algebras. This formally justifies and
extends the assertion of [11] concerning the
isomorphism of \ $\mathcal{A}$ \ and \ $\mathbb{R}$. \\

                                 \vspace{5mm}


\centerline{\sc 3. \ The \ $\stackrel{k}{\square}$ \ algebraic \
operations}

                                \vspace{3mm}

\noindent Instead of (3),(4) one could have initially made the
``reverse" identifications in (2.1),(2.2) of [11], namely
\begin{equation}
 x_{\{k\}} = e_k(x)
\end{equation}
and consequently
\begin{equation}
 x^{\{k\}} = \ln_k(x)
\end{equation}
With this identification, the requirements (2.3) and (2.4) of [11]
are satisfied if we define the following generalized addition
\begin{equation}
  x\stackrel{k}{\boxplus} y \ = \ \frac{\left\{ (1+kx)^\frac{1}{k}
          + (1+ky)^\frac{1}{k}\right\}^k - 1}{k}
\end{equation}
We notice that
\begin{displaymath}
 \lim_{k\rightarrow 0} \ x\stackrel{k}{\boxplus}y \ = \
 (e^x+e^y) \ln(e^x+e^y)
\end{displaymath}
 We easily check that \ $\stackrel{k}{\boxplus}$ \
is commutative, associative and has neutral element \
$-\frac{1}{k}$. \ As in the Section 3, the opposite of \ $x$, \
denoted by \ $\stackrel{k}{\boxminus}x$, \ is given by
\begin{equation}
  \stackrel{k}{\boxminus}x \ = \ \frac{(-1)^k (1+kx) - 1}{k}
\end{equation}

We can stop at this point to comment on the apparently
indiscriminate use of powers and logarithms throughout this work.
We have tacitly assumed that any real number can be raised to any
real power, or equivalently, that we can calculate the logarithm
of any non-zero real number. This is clearly possible, if we
consider the inclusion \ $j: \mathbb{R}\hookrightarrow
\mathbb{C}$. \ Then the logarithm \ $\ln z $ \ of \ $z=|z|
e^{i\theta} \in \mathbb{C}\backslash \{0 \}$, \ and the complex
power \ ($w\in\mathbb{C}$) \ are the (multi-valued) functions
\begin{displaymath}
  \ln z \ = \ \log |z| + i\theta  \hspace{2cm} z^w \ = \ e^{w \ln z}
\end{displaymath}
where \ $\log |z| $ \ stands for the usual logarithm of the
positive function of the modulus \ $|z|$. \ We choose,
arbitrarily, one branch of the logarithm \ $\ln$ \ and we work
with it everywhere [15]. Considering the reals as a subset of the
complex numbers is necessary, if we want to avoid complications
arising from the possible lack  of closure of the operations in
the sets of interest. A drawback of this approach is that such an
inclusion \ $j$ \ may obscure the direct physical interpretation
of some of the functions of physical interest.\\

Subtraction  is defined by \ $x\stackrel{k}{\boxminus}y =
x\stackrel{k}{\boxplus}(\stackrel{k}{\boxminus}y)$ \ and (25),(26)
give
\begin{equation}
 x\stackrel{k}{\boxminus} y \ = \ \frac{ \left\{(1+kx)^\frac{1}{k}
 + (-1)^k (1+ky)^\frac{1}{k} \right\}^k-1}{k}
\end{equation}
The generalized multiplication is defined by
\begin{equation}
 x\stackrel{k}{\boxtimes} y \ = \ x + y + kxy
\end{equation}
We notice that in the limit \ $k\rightarrow 0$, \ this operation
reduces to the usual addition in \ $\mathbb{R}$, \ namely
\begin{displaymath}
  \lim_{k\rightarrow 0} \ x\stackrel{k}{\boxtimes}y \ = \ x+y
\end{displaymath}
We can  check that \ $\stackrel{k}{\boxtimes}$ \ is commutative,
associative and has identity element \ $0$. \ In addition, we
observe that the distributivity property holds between \
$\stackrel{k}{\boxplus}$ \ and \ $\stackrel{k}{\boxtimes}$ \
namely
\begin{equation}
  x\stackrel{k}{\boxtimes}(y\stackrel{k}{\boxplus}z) \ = \
   (x\stackrel{k}{\boxtimes}y) \stackrel{k}{\boxplus}
   (x\stackrel{k}{\boxtimes}z)
\end{equation}
Therefore the structure $\mathcal{R}_2 = (\mathbb{R},
\stackrel{k}{\boxplus}, \stackrel{k}{\boxtimes})$ is a commutative
ring with identity. We can prove, by induction, that by
$\stackrel{k}{\boxplus}$-adding $n\in\mathbb{Z}_+$ times $x$,
\begin{equation}
x\stackrel{k}{\boxplus}\cdots \stackrel{k}{\boxplus}x \ = \
 \frac{ n^k(1+kx) - 1}{k}
\end{equation}
By using (30) we check that \ $\mathcal{R}_2$ \ has zero
characteristic. Motivated by (30), we define the multiplication \
$\stackrel{k}{\boxdot}$, \ for \ $n\in\mathbb{R}$
\begin{equation}
 n\stackrel{k}{\boxdot} x \ = \
 \frac{ n^k (1+kx) - 1}{k}
\end{equation}
We can also prove, by induction, that
$\stackrel{k}{\boxtimes}$-multiplying $x$ by itself
$n\in\mathbb{Z}_+$ times gives
\begin{equation}
 x\stackrel{k}{\boxtimes} \cdots \stackrel{k}{\boxtimes} x \ = \
  \frac{1}{k} \left\{ (1+kx)^n -1\right\}
\end{equation}
which can be used to prove that \ $\mathcal{R}_2$ \ has no
non-trivial nilpotent elements. The inverse element of \
$x\in\mathbb{R} \setminus \{ -\frac{1}{k} \}$, \ denoted by \
$\stackrel{k}{\boxslash}x$, \ is
\begin{equation}
  \stackrel{k}{\boxslash}x \ = \ \frac{-x}{1+kx}
\end{equation}
The division, as usual [11]-[13], is defined by \
$x\stackrel{k}{\boxslash}y =
x\stackrel{k}{\boxtimes}(\stackrel{k}{\boxslash}y)$ \ and we find,
by combining (28) and (33),
\begin{equation}
 x\stackrel{k}{\boxslash} y \ = \frac{x-y}{1+ky}
\end{equation}
So  \ $\mathcal{R}_2$ \ is actually a field. We can also check
that (2.6) and (2.7) of [11] become in this case
\begin{equation}
 \ln_k(x)\stackrel{k}{\boxplus}\ln_k(y) \ = \ \ln_k(x+y)
\end{equation}
\begin{equation}
 \ln_k(x)\stackrel{k}{\boxtimes}\ln_k(y) \ = \ \ln_k(xy)
\end{equation}
respectively. We proceed as in the construction of \ $\mathcal{A}$
in the previous paragraph. \ Using (31), we form an algebra \
$\mathcal{B}$ \ over \ $(\mathbb{R}, +,\cdot )$. Let \
$r,s\in\mathbb{R}$ \ and \ $x,y\in\mathcal{R}_2$. \ The following
relations, analogues of (17) - (20), are satisfied
\begin{equation}
 r\stackrel{k}{\boxdot} (x\stackrel{k}{\boxplus} y) \ = \
    (r\stackrel{k}{\boxdot}x) \stackrel{k}{\boxplus}
    (r\stackrel{k}{\boxdot}y)
\end{equation}
\begin{equation}
 (r+s) \stackrel{k}{\boxdot} x \ = \ (r\stackrel{k}{\boxdot} x)
 \stackrel{k}{\boxplus} (s\stackrel{k}{\boxdot} x)
\end{equation}
\begin{equation}
 r \stackrel{k}{\boxdot} (s\stackrel{k}{\boxdot} x) \ = \
  (rs) \stackrel{k}{\boxdot} x
\end{equation}
\begin{equation}
 1 \stackrel{k}{\boxdot} x \ = \ x
\end{equation}
Therefore, the structure \ $\mathcal{V}_2 = (\mathbb{R},
\stackrel{k}{\boxplus}, \stackrel{k}{\boxdot})$ \ is a vector
space over \ $(\mathbb{R},+,\cdot)$. \ The analogue of (21)
\begin{equation}
 r\stackrel{k}{\boxdot}(x\stackrel{k}{\boxtimes}y) \ = \
 (r\stackrel{k}{\boxdot}x)\stackrel{k}{\boxtimes}y \ = \
 x \stackrel{k}{\boxtimes} (r\stackrel{k}{\boxdot}y)
\end{equation}
is also satisfied. By the same arguments as for \ $\mathcal{A}$, \
one can prove that \ $\mathcal{B}$ \ is isomorphic to \
$\mathbb{R}$ \ or \
 $\mathbb{C}$ \ as division algebras.\\

It is worth mentioning that the matrix operations [16] over \
$\mathcal{R}_1$ \ and \ $\mathcal{R}_2$ \ are defined in the same
way as these over \ $\mathbb{R}$ \ if we replace $+$ and $\cdot$
with \ $\stackrel{k}{\oplus}$ \ and \ $\stackrel{k}{\otimes}$ \ or
with \ $\stackrel{k}{\boxplus}$ \ and \ $\stackrel{k}{\boxtimes}$.
\ Since there is no obvious and compelling reason to radically
change the definition of the Lie bracket between two matrices over
\ $\mathcal{R}_1$ \ or \ $\mathcal{R}_2$, \ we define it by,
respectively,
\begin{displaymath}
   [A,B]_{\tiny{\textcircled{k}}} \ = \
   \{ A\stackrel{k}{\otimes}B \} \stackrel{k}{\ominus}
   \{B \stackrel{k}{\otimes}A \}
            \hspace{2cm}
   [A,B]_{\tiny{\fbox{k}}} \ = \ \{A \stackrel{k}{\boxtimes}B \}
   \stackrel{k}{\boxminus} \{ B\stackrel{k}{\boxtimes}A \}
\end{displaymath}

                                \vspace{5mm}


\centerline{\sc 4. \ The \ $\textcircled{ k }$ and
$\tiny{\fbox{k}}$ \ differentials \ and \ integrals}

                                \vspace{3mm}

As in the case of algebraic operations, there is a lot of freedom
in defining  the \ $k$-deformed derivative [11], [13]. We work, in
most of this Section, with functions \ $f,g: \mathcal{R}_1
\rightarrow \mathcal{R}_1$ \ which will be differentiable as many
times as needed, as we usually do in Physics. Unlike previous
works [11]-[13], we believe that a more ``natural" definition of
the derivative \ $D_{\tiny{\textcircled{k}}}$ \ operator for such
functions  is
\begin{equation}
  D_{\tiny{\textcircled{k}}}f(x) = \lim_{y\rightarrow x}
  \{f(y)\stackrel{k}{\ominus} f(x)\} \stackrel{k}{\oslash}
  \{ y\stackrel{k}{\ominus} x\}
\end{equation}
By using (7) and (14) we find
\begin{equation}
 D_{\tiny{\textcircled{k}}}f(x) = \left\{ 1+ \frac{1}{x^{k-1}}\frac{d}{dx} [f(x)]^k
              \right\}^\frac{1}{k}
\end{equation}
We see that (43) gives
\begin{displaymath}
  \lim_{k\rightarrow 0} \ D_{\tiny{\textcircled{k}}} f(x) \ = \
  e^{x\frac{d}{dx} \ln f(x)}
\end{displaymath}
which is clearly not equal to \ $\frac{df}{dx}$. \ If \
$r\in\mathbb{R}$, \ we readily verify that
\begin{equation}
  D_{\tiny{\textcircled{k}}} \{ f(x) \stackrel{k}{\oplus} g(x) \} \ = \
          \{D_{\tiny{\textcircled{k}}}f(x)\}
          \stackrel{k}{\oplus} \{ D_{\tiny{\textcircled{k}}}g(x) \}
\end{equation}
\begin{equation}
  D_{\tiny{\textcircled{k}}} \{ r\stackrel{k}{\otimes} f(x) \} \ = \
  r \stackrel{k}{\otimes}\{ D_{\tiny{\textcircled{k}}}f(x) \}
\end{equation}
Therefore \ $ D_{\tiny{\textcircled{k}}}$ \ is a linear operator
with respect to \ $\stackrel{k}{\oplus}, \stackrel{k}{\otimes}$. \
We can also verify that Leibniz's rule holds
\begin{equation}
 D_{\tiny{\textcircled{k}}}\{f(x)\stackrel{k}{\otimes}g(x)\} \ = \
   \{ D_{\tiny{\textcircled{k}}}f(x) \}\stackrel{k}{\otimes} g(x)
   \stackrel{k}{\oplus}
   f(x) \stackrel{k}{\otimes} \{ D_{\tiny{\textcircled{k}}}g(x)\}
\end{equation}
Therefore \ $ D_{\tiny{\textcircled{k}}}$ \ is a derivation on the
space of functions \ $f: \mathcal{R}_1 \rightarrow \mathcal{R}_1.$
\ It is also interesting to notice that although
\begin{equation}
 D_{\tiny{\textcircled{k}}}\{r\stackrel{k}{\odot}f(x)\} \ = \
   r\stackrel{k}{\odot}\{ D_{\tiny{\textcircled{k}}}f(x) \}
\end{equation}
Leibniz's rule for \ $\stackrel{k}{\odot}$ \ does not hold, namely
\begin{equation}
D_{\tiny{\textcircled{k}}}\{f(x)\stackrel{k}{\odot}g(x)\} \ \neq \
   \{ D_{\tiny{\textcircled{k}}}f(x) \}\stackrel{k}{\odot} g(x)
   \stackrel{k}{\oplus}
   f(x) \stackrel{k}{\odot} \{ D_{\tiny{\textcircled{k}}}g(x)\}
\end{equation}
It is not difficult to find the reason for the failure of this
identity: \ $ D_{\tiny{\textcircled{k}}}$ \ is defined through \
$\stackrel{k}{\oslash}$ \ which is defined through \
$\stackrel{k}{\otimes}$, \ and the two ``circle" multiplications
$\stackrel{k}{\otimes}, \ \stackrel{k}{\odot}$ \ do not obey
obvious, ``nice" identities with each other. We notice that \ $
D_{\tiny{\textcircled{k}}}r=1, \ \forall \ r\in\mathcal{R}_1$ \
with $1$ being the neutral element with respect to \
$\stackrel{k}{\oplus}$ \ and \ $D_{\tiny{\textcircled{k}}}(x^k) =
(k+1)^\frac{1}{k}$ \ where the right-hand-side
is the identity with respect to \ $\stackrel{k}{\otimes}$.\\

Having defined the derivative operator \
$D_{\tiny{\textcircled{k}}}$ \ the next step is to find an
expression for the deformed exponential on \ $\mathcal{R}_1.$ \
Following [11], we demand the deformed exponential \
$\tilde{e}(x)$ \ to be an eigenfunction of \ $
D_{\tiny{\textcircled{k}}}$, \ namely to satisfy \
$D_{\tiny{\textcircled{k}}}\tilde{e}(x) = \tilde{e}(x).$ \ The
general solution of this differential equation is parametrized by
\ $c\in\mathbb{R}$ \ and is given by
\begin{equation}
  \tilde{e}_c(x) \ = \
  \left\{1+ce^\frac{x^k}{k}\right\}^\frac{1}{k}
\end{equation}
The deformed logarithm $\widetilde{\ln}_c(x)$, i.e. the inverse of
$\tilde{e}_c(x)$ with respect to composition, is
\begin{equation}
  \widetilde{\ln}_c(x) \ = \ \left\{
  k\ln\left(\frac{x^k-1}{c}\right)\right\}^\frac{1}{k}
\end{equation}
In the special case in which \ $c=k$, (49), (50) can be
re-expressed in terms of (3), (4) as
\begin{equation}
  \tilde{e}_k(x) \ = \ e_k\left(e^\frac{x^k}{k}\right)
\end{equation}
\begin{equation}
  \widetilde{\ln}_k(x) \ = \ k\ln(\ln_k(x))
\end{equation}
One can continue and define generalized hyperbolic and
trigonometric functions on \ $\mathcal{R}_1$ \ in terms of \
$\tilde{e}_k(x)$ \  and verify similar relations between them
[4]-[13]. Since this path is straightforward, we continue by
studying the integral operator. In the sequel will need the
differential
\begin{equation}
   d_{\tiny{\textcircled{k}}}x \ := \ \lim_{y\rightarrow x} \
    y\stackrel{k}{\ominus}x \
   = \  \left( 1+kx^{k-1}dx \right)^\frac{1}{k}
\end{equation}

The integral operator \ $\int_{\tiny{\textcircled{k}}}$ \ is
operationally defined as the inverse with respect to composition
of the differential operator \ $ D_{\tiny{\textcircled{k}}}$, \
namely, by demanding that
\begin{equation}
  D_{\tiny{\textcircled{k}}}\int_{\tiny{\textcircled{k}}}f(x)\stackrel{k}{\otimes}
  d_{\tiny{\textcircled{k}}}x \ = \ f(x)
\end{equation}
We can check that (54) is satisfied, if we define
\begin{equation}
 \int_{\tiny{\textcircled{k}}}f(x)\stackrel{k}{\otimes}d_{\tiny{\textcircled{k}}}x  \ := \
 \left\{ 1+ \int \left[(f(x))^k-1\right] x^{k-1} dx \right\}^\frac{1}{k}
\end{equation}
We immediately see that
\begin{displaymath}
  \lim_{k\rightarrow 0} \  \int_{\tiny{\textcircled{k}}}f(x)
     \stackrel{k}{\otimes}d_{\tiny{\textcircled{k}}}x \ = \
     e^{\int \ln f(x) dx}
\end{displaymath}
We can also verify that this integral operation is linear with
respect to \ $\stackrel{k}{\oplus}$ \ and \
$\stackrel{k}{\otimes}$, \ namely
\begin{equation}
  \int_{\tiny{\textcircled{k}}} \left\{ f(x)\stackrel{k}{\oplus}g(x)\right\}
  \stackrel{k}{\otimes} d_{\tiny{\textcircled{k}}}x \ = \
  \left\{ \int_{\tiny{\textcircled{k}}} f(x)
  \stackrel{k}{\otimes}d_{\tiny{\textcircled{k}}}x \right\}
  \stackrel{k}{\oplus} \left\{ \int_{\tiny{\textcircled{k}}}
  g(x)\stackrel{k}{\otimes}d_{\tiny{\textcircled{k}}}x \right\}
\end{equation}
and
\begin{equation}
 \int_{\tiny{\textcircled{k}}} \left\{ r\stackrel{k}{\otimes} f(x) \right\}
 \stackrel{k}{\otimes} d_{\tiny{\textcircled{k}}}x \ = \
 r\stackrel{k}{\otimes}\left\{ \int_{\tiny{\textcircled{k}}}
 f(x)\stackrel{k}{\otimes} d_{\tiny{\textcircled{k}}}x \right\}
\end{equation}

We can briefly turn our attention to the corresponding
constructions for the \ $\stackrel{k}{\square}$ \ operations. The
\ $D_{\tiny{\fbox{k}}}$ \ derivative is defined by analogy to (42)
as
\begin{equation}
   D_{\tiny{\fbox{k}}} f(x) \ = \  \lim_{y\rightarrow x}
   \{ f(y)\stackrel{k}{\boxminus} f(x) \} \stackrel{k}{\boxslash}
   \{ y\stackrel{k}{\boxminus} x\}
\end{equation}
A straightforward computation shows that (58) gives
\begin{equation}
 D_{\tiny{\fbox{k}}} f(x) \ = \left\{ \begin{array}{ll}
                                       \frac{f(x)-x}{1+kx}, \ &
                                       \mathrm{if} \ \
                                       k\neq 2m+1, \ m\in\mathbb{Z}\\
    \frac{1}{k} [ (\frac{df(x)}{dx})^k (\frac{1+kx}{1+kf(x)})^{k-1}
                     -1 ], \ & \mathrm{if} \ \ k = 2m+1, \ m\in\mathbb{Z}\\
                     \end{array} \right.
\end{equation}
We notice here that for any \ $k\in\mathbb{R}\setminus \{ 2m+1,
m\in\mathbb{Z}\}$, \ \ $D_{\tiny{\fbox{k}}}$ \ actually reduces to
a difference rather than an expression involving the derivative \
$\frac{df(x)}{dx}$. \ The case where \ $k$ \ is an odd integer is
the exception and has very little practical physical impact.
Indeed, one cannot determine the value of \ $k$ \ with infinite
precision for a system. When \ $k$ \ is an odd integer we can
always perturb it by adding an arbitrarily small decimal part to
it. We expect that the physical predictions arising from the
appropriate thermodynamic expressions should not be too sensitive
to such perturbations of \ $k$. \ Accordingly, the entropy for
such a small perturbation of \ $k$ \ should not change
dramatically. This requirement is a special case of the Lesche
stability criterion [17],[18] which is obeyed by the Tsallis'
entropy [2] as well as other forms of entropy [19],[20]. \\

Therefore, the case with \ $k$ \ being an odd integer effectively
never occurs in practice, so we do not have to elaborate on  a
mathematical structure describing it. For \
$k\in\mathbb{R}\setminus\{ \mathrm{Odd \ integer}\}$, \ we mention
for completeness that the ``integral" corresponding to \
$D_{\tiny{\fbox{k}}}$ \ is formally given by
\begin{equation}
  \int_{\tiny{\fbox{k}}} f(x)\stackrel{k}{\boxtimes}
  d_{\tiny{\fbox{k}}}x \ = \ x+(1+kx)f(x)
\end{equation}
We see that the \ $\stackrel{k}{\square}$ \ calculus operations do
not lead to anything particularly new and that is why we
do not pursue in this work any further properties they may satisfy.\\

                                \vspace{5mm}


\centerline{\sc 5. \ Discussion  \ and \ conclusions}

                                \vspace{5mm}

We defined the deformed operations \ $\stackrel{k}{\oplus}, \
\stackrel{k}{\otimes}, \ \stackrel{k}{\odot}, \
\stackrel{k}{\boxplus}, \ \stackrel{k}{\boxtimes}, \
\stackrel{k}{\boxdot}$. \ We have chosen not to modify the usual
composition of functions, since no such need arises, and
operations like matrix multiplication depend crucially on its
definition. \\

In previous works, either the distributivity property did not
hold, or the resulting structures were monoids and rigs at best.
With the operations that we defined,  the corresponding sets
become fields \ $\mathcal{R}_1$, \ $\mathcal{R}_2$, \ and
subsequently division algebras \ $\mathcal{A}$, \ $\mathcal{B}$ \
which turn out to be isomorphic to \ $\mathbb{R}$, \ $\mathbb{C}$.
\ We constructed a generalized derivative \
$D_{\tiny{\textcircled{k}}}$ \ and integral \ $
\int_{\tiny{\textcircled{k}}}$ \ on \ $\mathcal{R}_1$ and found
analogous, but not very interesting, structures on
$\mathcal{R}_2$. \ Most of these structures do not reduce to the
``usual" ones on \ $\mathbb{R}$ \ as \ $k\rightarrow 0$. \ The
unusual identification and the unexpected un-deformed limits of \
$\stackrel{k}{\bigcirc}$ \ and \ $\stackrel{k}{\square}$ \ can be
considered as being, partly, responsible for the fact that the
structures developed in this paper were previously missed
[12],[13]. The characterization as ``deformed" should be
understood to refer to the way the algebraic and calculus
operations were generated, rather than to their actual \
$k\rightarrow 0$ \ limit. Although the above structures are
mathematically attractive, their physical relevance in quantifying
ideas of non-extensive thermo-statistics
is not clear at this point.\\

We have performed a very heuristic, operational, treatment  of the
concepts of limit, derivative and integral in \ $\mathcal{R}_1$ \
and \ $\mathcal{R}_2$. \ A more careful approach may probably be
warranted in the future, but our treatment seems to be adequate
for present purposes. Someone can extend our results by
constructing more elaborate algebraic structures, like bi-algebras
and Hopf algebras or can follow the steps of homological algebra
[21]  and proceed by constructing differential complexes [22] etc.
Apart from some potential mathematical significance, it is not
clear to us what the use of such structures would be in
non-extensive thermo-statistics, or in any other context, that is
why we did not
pursue their further development in this work.\\

                                 \vspace{2mm}


\noindent {\sc Acknowledgement:} \ We are grateful to Professor G.
Kaniadakis for his comments on the manuscript. We would like to
thank Professor E.P. Borges for sending us a copy of [4].\\



                                  \newpage

\centerline{\normalsize\sc References}

                                \vspace{2mm}

\setlength{\baselineskip}{24pt}
\noindent\rm 1. C. Tsallis, \emph{J. Stat. Phys.} {\bf 52,} 479 (1988).\\
2. \emph{Nonextensive entropy; Interdisciplinary Applications},
     C. Tsallis, M.Gell-Mann, Eds., \\  \hspace*{4.5mm} Oxford Univ. Press (2004).\\
3. C. Tsallis, \ {\sf arXiv:cond-mat/0409631}\\
4. E.P. Borges, \emph{J. Phys. A} {\bf 31,} 5281 (1998).\\
5. C. Tsallis, \emph{Braz. J. Phys.} {\bf 29,} 1 (1999).\\
6. E.K. Lenzi, E.P. Borges, R.S. Mendes, \emph{J. Phys. A}
   {\bf 32,} 8551 (1999).\\
7. G. Kaniadakis, \emph{Physica A} {\bf 296,} 405 (2001).\\
8. G. Kaniadakis, A.M. Scarfone, \emph{Physica A} {\bf 305,} 69 (2002).\\
9. T. Yamano, \emph{Physica A} {\bf 305,} 486 (2002).\\
10. J. Naudts, \emph{Physica A} {\bf 316,} 323 (2002).\\
11. G. Kaniadakis, \emph{Phys. Rev. E} {\bf 66,} 056125 (2002). \\
12. L. Nivanen, A. Le M\'{e}haut\'{e}, Q.A. Wang, \emph{Rep.
      Math. Phys.} {\bf 52,} 437 (2003).\\
13. E.P. Borges, \emph{Physica A} {\bf 340,} 95 (2004).\\
14. T.W. Hungerford, \emph{Algebra}, Springer (1980).\\
15. A.I. Markushevich, \emph{Theory of Functions of a Complex
Variable}, 2nd Ed.,   Chelsea  (1977).\\
16. W. Greub, \emph{Linear Algebra}, 4th Ed., Springer (1981).\\
17. B. Lesche, \emph{J. Stat. Phys} {\bf 27,} 419 (1982).\\
18. S. Abe, \emph{Phys. Rev. E} {\bf 224,} 046134 (2002).\\
19. G. Kaniadakis, A.M. Scarfone, \emph{Physica A} {\bf 340,} 102 (2004).\\
20. S. Abe, G. Kaniadakis, A. M. Scarfone, \ {\sf arXiv:cond-mat/0401290}\\
21. P.J. Hilton, U. Stammbach, \emph{A course in Homological
    Algebra}, 2nd Ed., Springer (1997).\\
22. D. Eisenbud, \emph{Commutative Algebra with a View Toward
    Algebraic Geometry},\\ \hspace*{5.5mm} Springer (1995).\\

                           \vfill

\end{document}